\documentclass[a4paper,12pt]{article}
\pdfoutput=1
\usepackage{feynmp-auto,expdlist}
\usepackage{amsmath, amsfonts, amssymb}
\usepackage{graphicx}
\usepackage{enumerate}
\usepackage{hyperref}
\usepackage{latexsym}
\usepackage{dsfont}
\usepackage{hepnicenames}
\usepackage{enumerate}
\usepackage{soul}
\usepackage[normalem]{ulem}
\usepackage{bbold}

\usepackage{units}

\usepackage{mathrsfs,graphicx,rotating,amsmath,amsfonts,mathtools,booktabs,amssymb,wasysym}
\usepackage{hyperref}\usepackage{slashed}
\usepackage[nosort]{cite}
\usepackage[table,xcdraw,dvipsnames]{xcolor}
\usepackage{bm}
\usepackage{graphicx}
\usepackage{multirow,multicol}
\hypersetup{colorlinks,bookmarksopen,bookmarksnumbered,
linkcolor=blus,pdfstartview=FitH,urlcolor=rossos,citecolor=verde}
\allowdisplaybreaks

\renewcommand{\[}{\left[}

\newcommand{\mio}[1]{}

 \newcommand{\med}[1]{\langle #1\rangle}

\def\bpm{\begin{pmatrix}}
\def\epm{\end{pmatrix}}

\usepackage{mathrsfs}

 \newcommand{\fig}[1]{~\ref{fig:#1}}
\newcommand{\sfrac}[2]{#1/#2}

\allowdisplaybreaks
\usepackage{multicol}
\usepackage{color}
\definecolor{rosso}{cmyk}{0,1,1,0.4}
\definecolor{rossos}{cmyk}{0,1,1,0.55}
\definecolor{rossoc}{cmyk}{0,1,1,0.2}
\definecolor{blu}{cmyk}{1,1,0,0.3}
\definecolor{blus}{cmyk}{1,1,0,0.6}
\definecolor{bluc}{cmyk}{1,1,0,0.1}
\definecolor{verde}{cmyk}{0.92,0,0.59,0.25}
\definecolor{verdec}{cmyk}{0.92,0,0.59,0.15}
\definecolor{verdes}{cmyk}{0.92,0,0.59,0.4}

\newcommand{\eq}[1]{~{\rm (\ref{eq:#1})}}

\newcommand{\GeV}{\,{\rm GeV}}
\newcommand{\TeV}{\,{\rm TeV}}

\def\circa#1{\,\raise.3ex\hbox{$#1$\kern-.75em\lower1ex\hbox{$\sim$}}\,}

\newcommand{\beq}{\begin{equation}}
\newcommand{\eeq}{\end{equation}}

\newcommand{\bea}{\begin{eqnarray}}
\newcommand{\eea}{\end{eqnarray}}
\newcommand{\be}{\begin{equation}}
\newcommand{\ee}{\end{equation}}
\font\tenrsfs=rsfs10 at 12pt
\font\sevenrsfs=rsfs7
\font\fiversfs=rsfs5
\newfam\rsfsfam
\textfont\rsfsfam=\tenrsfs
\scriptfont\rsfsfam=\sevenrsfs
\scriptscriptfont\rsfsfam=\fiversfs

\newsavebox\MBox

\def\circa#1{\,\raise.3ex\hbox{$#1$\kern-.75em\lower1ex\hbox{$\sim$}}\,}
\makeatletter

\font\ital=cmu10

\def\hhref#1{\href{http://arxiv.org/abs/#1}{arXiv:#1}}
\usepackage{xstring}
\newcommand{\hhrefq}[1]{\IfSubStr{#1}{:}{\href{http://inspirehep.net/search?ln=en&ln=en&p=#1&of=hb&action_search=Search&sf=&so=d&rm=&rg=25&sc=0}{InSpire:#1}}{\hhref{#1}}}

\def\art{\@ifnextchar[{\eart}{\oart}}
\def\eart[#1]#2#3#4#5#6{{\rm #2}, {\em #3 \bf #4} {\rm (#6) #5} ({\em #1})}
\def\article{\@ifnextchar[{\earticle}{\oarticle}}
\def\oarticle#1#2#3#4#5#6{{\rm #1}, {\ital ``#6''}, {\rm #2 #3 (#5) #4}}
\def\earticle[#1]#2#3#4#5#6#7{{\rm #2}, {\ital ``#7''}, {\rm #3 #4 (#6) #5}  [\hhrefq{#1}]}
\def\hepart[#1]#2{{\rm #2, \sl#1}}
\def\heparticle[#1]#2#3{#2, {\ital ``#3''} [\hhrefq{#1}]}
\newcommand{\doi}[1]{\href{http://dx.doi.org/#1}{[link]}}

\newcommand{\hhrefqq}[1]{\IfBeginWith{#1}{10.}{\href{https://doi.org/#1}{doi:#1}}{\hhrefq{#1}}}
\def\earticle[#1]#2#3#4#5#6#7{{\rm #2}, {\ital ``#7''}, {\rm #3 #4 (#6) #5}  [\hhrefqq{#1}]}

\renewenvironment{thebibliography}[1]
     {\begin{multicols}{2}[\section*{\refname}]%
      \@mkboth{\MakeUppercase\refname}{\MakeUppercase\refname}%
      \list{\@biblabel{\@arabic\c@enumiv}}%
           {\settowidth\labelwidth{\@biblabel{#1}}%
            \leftmargin\labelwidth
            \advance\leftmargin\labelsep
            \@openbib@code
            \usecounter{enumiv}%
            \let\p@enumiv\@empty
            \renewcommand\theenumiv{\@arabic\c@enumiv}}%
      \sloppy
      \clubpenalty4000
      \@clubpenalty \clubpenalty
      \widowpenalty4000%
      \sfcode`\.\@m}
     {\def\@noitemerr
       {\@latex@warning{Empty `thebibliography' environment}}%
      \endlist\end{multicols}}

\newcounter{alphaequation}[equation]
\def\thealphaequation{\theequation\hbox to
0.6em{\hfil\alph{alphaequation}\hfil}}
\def\eqnsystem#1{
\def\@eqnnum{{\rm (\thealphaequation)}}
\def\@@eqncr{\let\@tempa\relax \ifcase\@eqcnt \def\@tempa{& & &} \or
  \def\@tempa{& &}\or \def\@tempa{&}\fi\@tempa
  \if@eqnsw\@eqnnum\refstepcounter{alphaequation}\fi
\global\@eqnswtrue\global\@eqcnt=0\cr}
\refstepcounter{equation} \let\@currentlabel\theequation \def\@tempb{#1}
\ifx\@tempb\empty\else\label{#1}\fi
\refstepcounter{alphaequation}
\let\@currentlabel\thealphaequation
\global\@eqnswtrue\global\@eqcnt=0 \tabskip\@centering\let\\=\@eqncr
$$\halign to \displaywidth\bgroup \@eqnsel\hskip\@centering
$\displaystyle\tabskip\z@{##}$&\global\@eqcnt\@ne
\hskip2\arraycolsep\hfil${##}$\hfil& \global\@eqcnt\tw@\hskip2\arraycolsep
$\displaystyle\tabskip\z@{##}$\hfil
\tabskip\@centering&\llap{##}\tabskip\z@\cr}
\def\endeqnsystem{\@@eqncr\egroup$$\global\@ignoretrue} \makeatother

\definecolor{Gray}{gray}{0.95}

\def\bal#1\eal{\begin{align}#1\end{align}}

\newcommand{\DVmax}{\Delta V_{\rm max}}

\newcommand{\Hinf}{H_{\rm infl}}

\begin{document}
\vspace{1.5cm}

\begin{center}
{\Large\LARGE \bf \color{rossos}
Relaxing the Higgs mass\\ and its vacuum energy\\ by living at the top of the potential\\[0.8em]}
{\bf Alessandro Strumia$^{a}$, Daniele Teresi$^{a,b}$}\\[7mm]

{\it $^a$ Dipartimento di Fisica dell'Universit{\`a} di Pisa}\\[1mm]
{\it $^b$ INFN, Sezione di Pisa, Italy}\\[1mm]

\vspace{0.5cm}

\begin{quote}\large\color{blus} 
We consider an ultra-light scalar coupled to the Higgs in the presence of heavier new physics.
In the electroweak broken phase the Higgs gives a tree-level contribution to the light-scalar potential,
while new physics contributes at loop level.
Thereby, the theory has a cosmologically meta-stable 
phase where the light scalar is around the top of its potential, 
and the Higgs is a loop factor lighter than new physics.
Such regions with precarious naturalness are anthropically and environmentally selected,
as regions with heavier Higgs crunch quickly.
We expect observable effects of rolling in the dark-energy equation of state.
Furthermore, vacuum energies up to the weak scale can be canceled down
to anthropically small values. 
\end{quote}

\thispagestyle{empty}
\bigskip

\end{center}

\setcounter{footnote}{0}

\setcounter{tocdepth}{1}
\tableofcontents

\section{Introduction}
Scalars with mass $m$ below the present Hubble scale $H_0$ can still lie away from the minimum of their potential
and thereby undergo significant cosmological evolution at present times.
This can  be of possible relevance for understanding the apparently unnatural hierarchy of scales observed in Nature:
cosmological constant much below the weak scale
much below the Planck scale.
See~\cite{Dvali:2003br,Dvali:2004tma,Graham:2015cka,Arkani-Hamed:2016rle,Arvanitaki:2016xds,Alberte:2016izw,Geller:2018xvz,Cheung:2018xnu,Graham:2019bfu,Strumia:2019kxg,Giudice:2019iwl,Bloch:2019bvc} for some recent attempts of understanding hierarchies via cosmological evolution.

\smallskip

We here consider a scalar $\phi$ with the shift symmetry that keeps its potential flat
broken by small interactions to other fields, in particular to the Higgs
doublet $H$ and to other heavier new physics that generates a Higgs mass hierarchy problem.
The interaction with the Higgs can be parametrized by a $\phi$-dependent Higgs mass,
$M_h^2(\phi)$.
The model will cosmologically generate a little hierarchy, maximal in the special case
where the $\phi$ couplings to new physics are mediated by the Higgs.

Indeed, integrating out the Higgs generates a {\em tree-level} negative
contribution $V_\phi \sim - M_h^4(\phi)$ to the effective potential for $\phi$
if $M_h^2(\phi)>0$, i.e.~when electroweak symmetry breaking takes place.
Heavier new physics contributes at {\em loop level}.
The full potential $V_\phi$ can have a {\em maximum} for values of $\phi$
such that the Higgs mass is a loop factor lighter than new physics.


This observation is relevant for the Higgs-mass hierarchy problem if,
for some reason, $\phi$ lies close to the maximum of its potential.
The authors of~\cite{Geller:2018xvz,Cheung:2018xnu} explored possible reasons:
barring the possibility that $\phi$ is a ghost, they added
extra wiggles to $V_\phi$ in order to generate local minima around the maximum,
arguing that they can be favored cosmologically because they inflate more.



\begin{figure}
$$\includegraphics[width=0.5\textwidth]{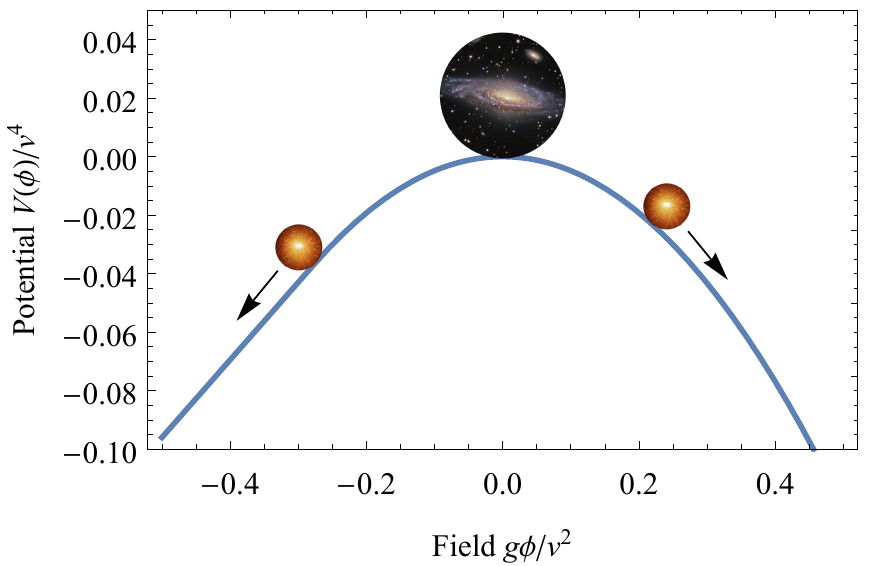}$$
\caption{\it  Effective potential for $\phi$ around its maximum. 
After inflation different patches of the Universe have different values of $\phi$. 
In most patches $\phi$ is away from its maximum and slides down leading to a crunch.
Only patches where $\phi$ is near the top are meta-stable on cosmological time-scales.
These correspond to a Higgs mass one or two loop factors lighter than new physics. \label{fig:potential}}
\end{figure}

We consider a simpler way of living at the top. 
Since the potential $V_\phi$ is flat around its maximum, regions sufficiently close to it 
can be cosmologically meta-stable, as long as $\phi$ is so weakly coupled that its mass is small, $m\circa{<}H_0$. 
Instead, regions away from the maximum quickly roll down
and catastrophically collapse into a big crunch, as illustrated in fig.\fig{potential}.
In this way, only regions where the Higgs is light are long-lived: a small Higgs mass is selected.
The multiverse volume at late times gets dominated by such near-critical regions 
at the border between the broken and unbroken Higgs phase.

We also find that this setting allows for a range of values of the vacuum energy,
up to the weak scale in the most optimistic case,
thereby providing a mechanism (alternative to a landscape of vacua)
for the anthropic selection of the cosmological constant down to the small observed value.


This scenario generates a little hierarchy, making the weak scale
one or two loop factors below the scale of  new physics.
The needed new physics is unspecified and might be a full solution to hierarchy problems, such as supersymmetry
(see e.g.~\cite{Cheung:2018xnu}),
or possibly a landscape with electroweak, rather than meV, spacing (e.g.~from string~\cite{Bousso:2000xa,Kachru:2003aw,Susskind:2003kw} or quantum field~\cite{ArkaniHamed:2005yv,Ghorbani:2019zic} theory).

An observable consequence of  \emph{precarious naturalness} is that in an $\mathcal{O}(1)$ fraction of the long-lived regions of the Universe, the scalar field $\phi$ is rolling down the potential \emph{now}, 
with detectably large effects on the equation of state of dark energy. 
The other robust consequence is the presence of new physics coupled to the Higgs
around or below the $\approx 20 \TeV$ scale, in order to generate the maximum at the observed value. 

\smallskip

The paper is structured as follows.
In section~\ref{model} we present the model and  study its cosmological dynamics in section~\ref{sec:detailsmechanism}, where we illustrate the mechanism leading to precarious naturalness. We then discuss bounds and signals in section~\ref{sec:bounds} and finally conclude in section~\ref{sec:conclusions}.


\section{Precarious naturalness: the setup}\label{model}
We consider the Standard Model (SM) with mass scale $M_h \approx 125\GeV$
plus heavier new physics at the mass scale $M \gg M_h$ and
an ultra-light scalar $\phi$ with mass-scale $m \ll M_h$.

A Higgs-mass hierarchy problem arises if
the Higgs interacts significantly with heavier new particles.
We leave the heavy new physics unspecified.
This includes the possibility of new physics, such as supersymmetry, that makes
its scale $M$ naturally smaller than the ultimate cut-off of the theory at some scale
$\Lambda$, possibly around the Planck mass.\footnote{The possibility of new physics at the weak scale that makes the Higgs mass natural would have been simpler, but it has been disfavored by collider searches.}

On the other hand, the lightness of $\phi$ is naturally understood if the $\phi$ potential is protected by a shift symmetry,
broken at some scale $\circa{<}\Lambda$ by very small couplings $g$. \footnote{Such tiny couplings mediate a force much weaker than gravity for the Higgs. However, this is not in contradiction with the Weak Gravity Conjecture~\cite{ArkaniHamed:2006dz}, since the force is attractive, so that its effect adds up to gravity~\cite{1705.04328}. Sometimes it is argued that also scalar forces must be stronger than gravity~\cite{1705.04328}, but the arguments for this are  weaker~\cite{1903.06239} and disappear for fields with sizable gauge interactions, like the Higgs boson in our case.}
For example, the heavy new physics could be supersymmetric and $\phi$ could be a quasi-modulus.

We want to compute the effective potential for $\phi$.
First, we integrate out the heavy new physics obtaining the 
effective field theory around the weak scale, with effective potential
\beq \label{eq:pot_tree}
 V_{\rm eff}(H,\phi) = V_0 - \frac{M_h^2(\phi)}{2} |H|^2 + \lambda |H|^4 + V^{\rm heavy}_\phi (\phi) ,\eeq
where $\lambda \approx 0.126$ is the Higgs quartic.
The potential contains the tree-level coupling of $\phi$ to the Higgs, parametrized by $M_h^2(\phi)$.
The squared Higgs mass might receive unnaturally large corrections from heavy
new physics, but what matters here is its value renormalized around the weak scale.
In agreement with the decoupling theorem, 
heavy new physics can instead give the $V^{\rm heavy}_\phi (\phi)$ term in the effective potential
(to be computed at one- or two-loop level in section~\ref{Vheavy}).

Finally, we integrate out the Higgs boson. 
If $M_h^2(\phi) >0$ (such that, in our notation, the electroweak symmetry is broken),
the Higgs gives a tree-level contribution to $V_{\rm eff}(\phi)$.
The Higgs also gives a loop-level contribution,
the usual running of the cosmological constant,
that can be neglected with respect to loop effects of heavier fields.
We thereby obtain the effective potential for the light field $\phi$
\beq \label{eq:pot_loop}
V_{\rm eff}(\phi)  = V_0 - \frac{M_h^4(\phi)}{16\lambda}  \theta(M_h^2(\phi))+ V^{\rm heavy}_\phi (\phi) , \eeq
where $\theta=1$ for $M_h^2(\phi)>0$ and zero otherwise, 
so that the tree-level part of $V_{\rm eff}(\phi)$ is flat for $\phi<0$.

We assume that the sign of $V_\phi^{\rm heavy} $ is such that, together with the negative tree-level Higgs term,
 $V_{\rm eff}(\phi)$ has a local maximum.
Without loss of generality, we can shift $\phi$ so that the maximum lies at $\phi=0$.
We next assume, for simplicity, that the two terms can be approximated by first-order Taylor expansions
\beq M_h^2 (\phi) \simeq  M_{h0}^2 + g \phi,\qquad
V_\phi^{\rm heavy} \simeq g A \phi .
\eeq
The parameter $M_{h0}$ must be negligibly different from the physical Higgs mass, $125\GeV$,
given that  $\phi$ will lie very close to the maximum.
The maximum condition then implies $M_{h0}^2=8 \lambda A$.
As we now show, this is loop suppressed  with respect to $M$.


\subsection{The loop contribution to $V_\phi$ from heavy new physics}\label{Vheavy}
If the full theory is valid up to some scale $\Lambda\gg M$, 
a dominant contribution to $V_\phi^{\rm heavy}$
is computed by solving the one-loop RGE in the full theory.
Otherwise $V_\phi^{\rm heavy}$ arises as threshold corrections at the scale $M \approx \Lambda$.
For the sake of illustration, we compute the minimal contribution to $V_\phi^{\rm heavy}$ in a 
simple toy model:
we assume that the new physics at scale $M$ is just a singlet scalar $S$ with quartic coupling $\lambda_{HS} S^2 |H|^2$ to the Higgs.

\bigskip

To start,  in order to compute the minimal $V_\phi^{\rm heavy}$ at two loops,
we assume that at some UV scale $\Lambda \circa{>}M$ the shift symmetry of $\phi$ is broken only by its coupling to
the Higgs. 
RGE running at one loop in the full theory above $M$ contains the following two terms:
\beq
\frac{d M_h^2(\phi)}{d\ln\mu} =  - \frac{4 \lambda_{HS}^2 M^2}{(4\pi)^2} ,\qquad
\frac{d V_\phi}{d\ln\mu} =  \frac{M_h^4(\phi)}{2 (4\pi)^2} +\cdots.
\eeq
In words: heavy new physics contributes to the Higgs mass
that contributes to the RGE running of the cosmological constant.
As $\phi$ can be treated as a background field, in view of its small coupling,
we obtain the $\phi$ potential by iteratively solving the two RGE equations.
Apart from a $\phi$-independent constant, the dominant term is
\be 
V_\phi^{\rm heavy}(\phi) \simeq  \frac{M^2 M_h^2(\phi) }{(4 \pi)^4} \, 2 \lambda_{HS}^2 \ln^2\!\frac{\Lambda}{M} \;,
\ee
under the assumption that the RGE correction to $M_h^2$ from new physics is much larger than $M_h^2$ itself
(otherwise, no Higgs mass naturalness problem is present~\cite{1303.7244}).
This RGE effect from boson loops has the desired sign\footnote{In order to see this, note that the boundary conditions for the running are $V_{\phi}^{\rm heavy}|_{\mu = \Lambda} = 0$, $M_h^2(\phi)|_{\mu \simeq M} = M_h^2(\phi)$. The former because at the scale $\Lambda$ the shift symmetry is assumed to be broken only by the coupling to the Higgs, the latter because the potential minimization is performed at the scale $v \approx M$.}.
In this more favorable scenario $V_\phi^{\rm heavy}$ is two-loop suppressed,
so that it has the size needed by the mechanism under study
for 
\beq \label{eq:2loop}M \sim (4\pi)^2 M_h \sim20\TeV,\eeq
having assumed $\ln \Lambda/M\sim 1$ and $\lambda_{HS}\sim 1$.
Order one couplings are unavoidably needed
in models where
new physics at mass $M$ provides a natural solution to the big hierarchy problem
(for example in supersymmetry the Higgs has top Yukawa couplings to sparticles).

\bigskip

The less favorable possibility is that both $H$ and new physics $S$ directly couple to $\phi$.
We can parametrize the new physics direct coupling as $M_S(\phi)^2 \simeq M^2 - g_S \phi$, with the shift-symmetry breaking coupling $g_S \sim g$. RGE running of $V_\phi$ above $M$ contains the term
$\sfrac{d V_\phi}{d\ln\mu} =  \sfrac{M_S^4(\phi)}{2 (4\pi)^2} +\cdots$,
giving rise to a one-loop suppressed $V_\phi^{\rm heavy}$:
\be 
V_\phi^{\rm heavy}(\phi) \simeq  \frac{M^2 g_S \phi }{(4 \pi)^2} \ln\!\frac{\Lambda}{M} \;.
\ee
Assuming  $\ln \Lambda/M\sim 1$, the scale of new physics
needed by the mechanism now is
\beq \label{eq:1loop}M \sim 4\pi M_h \sim 1.5 \TeV.\eeq

%
%

%


%
%
%

\section{Precarious naturalness: cosmological dynamics}\label{sec:detailsmechanism}
We here outline the main cosmological dynamics that results in precarious naturalness.
Possible issues will be addressed in the next section, finding bounds that can be satisfied.

In the primordial Universe,  inflation creates patches with different values of $\phi$. 
The small $V_{\rm eff}(\phi)$ starts playing a role only in the late Universe. 
In most patches $\phi$ is away from the maximum and
slides down on a time scale much smaller than the cosmological one, $1/H_0$. 
What is inside these patches undergoes a big crunch.
Only in rare regions the field $\phi$ sits close enough to its maximum giving rise to
a meta-stable Universe.
These regions, characterized by a small Higgs mass,
get selected both anthropically and environmentally.
Environmentally because the multi-verse volume, at late times, gets dominated
by meta-stable regions that do not crunch quickly
 (see also~\cite{Strumia:2019kxg}).\footnote{Qualitatively different regions with a positive vacuum energy $V_0$ so large that they survive to $\phi$ sliding down 
are excluded only anthropically because of the large cosmological constant.
Volumes can be estimated as
$$
\mathcal{V}^{(3)}_{\rm collapsed} \lesssim \frac{M_{\rm Pl}^3}{M^6},\qquad
\mathcal{V}^{(3)}_{\rm meta-stable}  \gtrsim \frac{1}{H_0^3} \frac{M_{\rm Pl} H_0^2}{m \Lambda^2} ,
$$
as we now explain.
Since collapsed patches are $\sim$ AdS we need to specify a measure: 
we calculate the volume as seen from inside~\cite{Coleman:1980aw,Espinosa:2015qea}, pessimistically assuming
that the bottom of the potential is at $V \sim - M^4$ (more presumably is $V \sim - \Lambda^4$). 
The volume of meta-stable patches is conservatively taken as $1/H_0^3$ times
the small probability $\phi_{\rm max}/\Delta \phi_{\rm tot}$, where the latter is the total  range of $\phi$. 
We pessimistically assume $g \Delta \phi_{\rm tot} \sim \Lambda^2$. 
Then, the volume with light Higgs certainly dominates if
$ \Lambda <  \sfrac{M^3}{H_0 M_{\rm Pl}}$, which is much above the Planck scale. 
A fortiori, the 4-volume of the meta-stable regions dominates the 4-volume of the Universe.}
Anthropically because observers can form only in these long-lived regions.

The classical cosmological evolution of a homogeneous scalar $\phi$ 
around the top of its potential $V(\phi) = - m^2\phi^2/2$, with $m = g/\sqrt{8 \lambda}$, 
is described by
\be\label{eq:EoMoriginal}
\frac{d^2 \phi}{d t^2} + 3 H \frac{d \phi}{d t} = m^2 \phi ,
\ee
where $H = \dot a/a$ is the Hubble rate. As long as $H\gg m$ the field slow rolls as
\beq \phi (t) \approx \phi(t_i)\exp\left(\int_{t_i}^t \frac{m^2}{3H^2} \, d\ln a\right).\eeq
So, after the time $t_*$ when $H(t_*)$ becomes smaller than $m$,
the field grows exponentially with time-scale $m$:
$\phi(t)\approx \phi(t_*) e^{m (t-t_*)}$.
Even for $m \ll H_0$, the slower rolling down leads to a deadly big crunch before the present epoch
if the induced variation in $V_\phi$ is larger than the present vacuum energy.

In order to compute the allowed range of $\phi$, we impose that the change in $V(\phi)$ up to now
(Hubble constant $H_0$) is smaller than some $\DVmax $.
Neglecting ${\cal O}(1)$ factors, this
restricts the initial value of the field to be close to the maximum within
\be\label{eq:phimax}
|\phi| \circa{<} \phi_{\rm max} \equiv \sqrt{\DVmax} 
\left\{\begin{array}{ll}
  {H_0}/m^2    & \hbox{if }m \lesssim H_0 \\
  e^{- m/H_0} /{m}  & \hbox{if } m \gtrsim H_0
    \end{array}\right. .
\ee
The condition that cosmological evolution leads to an anthropically acceptable 
long-lived Universe with age $\approx 1/H_0$ 
corresponds to
$\DVmax \sim H_0^2 M_{\rm Pl}^2$, so that
\beq \label{eq:phirange}
\phi_{\rm max}\approx
M_{\rm Pl}
 \left\{\begin{array}{ll}
  {H_0^2}/m^2    & \hbox{if }m \lesssim H_0 \\
  e^{- m/H_0} H_0/{m}  & \hbox{if } m \gtrsim H_0
    \end{array}\right. .
\eeq
The range $|\phi|\circa{<}\phi_{\rm max}$ corresponds to a range of Higgs masses
\beq |M_h^2 - M_{h0}^2| \circa{<} g|\phi| \sim g M_{\rm Pl}  
 \left\{\begin{array}{ll}
  {H_0^2}/m^2    & \hbox{if }m \lesssim H_0 \\
  e^{- m/H_0} H_0/{m}  & \hbox{if } m \gtrsim H_0
    \end{array}\right. ,
\eeq
which is not larger than the observed small weak scale $v$ if
\beq \label{eq:HiggsNat}
m \circa{>} H_0^2 M_{\rm Pl}/v^2 \sim 10^{-29} H_0.\eeq
Eq.s\eq{phimax} and \eq{phirange} give the maximal range of $\phi$ such that its time evolution up to now
leads to a change in vacuum energy $V_\phi(\phi) = -m^2 \phi^2/2$ smaller than the observed vacuum energy
$\Delta V_{\rm max}$.
This field range contains a range $V_{\rm max}$ of vacuum energies 
larger than the time variation $\Delta V_{\rm max}$ if $m \circa{<}H_0$:
\beq V_{\rm max} = \Delta V_{\rm max} \left(\frac{H_0}{m}\right )^2 \circa{<} v^4  .\eeq
Inserting $\Delta V_{\rm max} \sim H_0^2 M_{\rm Pl}^2$,
the maximal value of $V_{\rm max}$ is achieved assuming the far-end lower range for $m$ in
eq.\eq{HiggsNat}.
Such ultra-ultra-light scalar allows also to tune vacuum energies
as large as the weak scale $v^4$ down to  acceptable small values,
of order $H_0^2 M_{\rm Pl}^2$, 
as demanded by anthropic considerations
(see \cite{Weinberg:1988cp} and subsequent works)
and provides a physical mechanism to realize such anthropic selection.
The situation is illustrated in fig.\fig{CC}. 
If the unknown constant $V_0$ is smaller than about $v^4$,
some field value within the range $|\phi| < \phi_{\rm max}$, allowed by stability up to now,
contains the observed small vacuum energy, and is realized in some long-lived patch (left panel).
Larger values of $V_0$ only allow short-lived patches that roll down cosmologically fast (right panel),
strengthening the usual argument about anthropic selection of the cosmological constant~\cite{Weinberg:1988cp}.

%

\section{Bounds and signals}\label{sec:bounds}
We next consider various bounds, showing that $\phi$ must be lighter than $\sim 15 H_0$.
With this extra condition, the above discussion survives to all constraints. Then, we show that a generic prediction of our framework is the presence of sizable effects on the dark-energy equation of state.

\subsection{Bound from quantum fluctuation}

We need to impose that quantum fluctuations
($\delta \phi \approx m$ per time $1/m$)
and de Sitter fluctuations ($\delta \phi \sim H_0$ per Hubble time)
leave the field $\phi$ within the meta-stability range,
$|\phi| \circa{<}\phi_{\rm max}$ as computed in eq.~\eqref{eq:phimax}. 
For $m \circa{<} H_0$ the condition is satisfied.
For $m \circa{>} H_0$ the desired range is much smaller because
the  instability in $V_\phi$ amplifies fluctuations 
by the $e^{m t}\sim e^{m/H_0}$ factor, making them classical (see e.g.~\cite{Guth:1985ya}).
The field remains within the allowed range if it is not much heavier than the Hubble scale
\beq m \circa{<} H_0 \ln \frac{M_{\rm Pl}}{H_0} \sim 100 H_0 \,.\eeq

%

\begin{figure}
$$\includegraphics[width=0.4\textwidth]{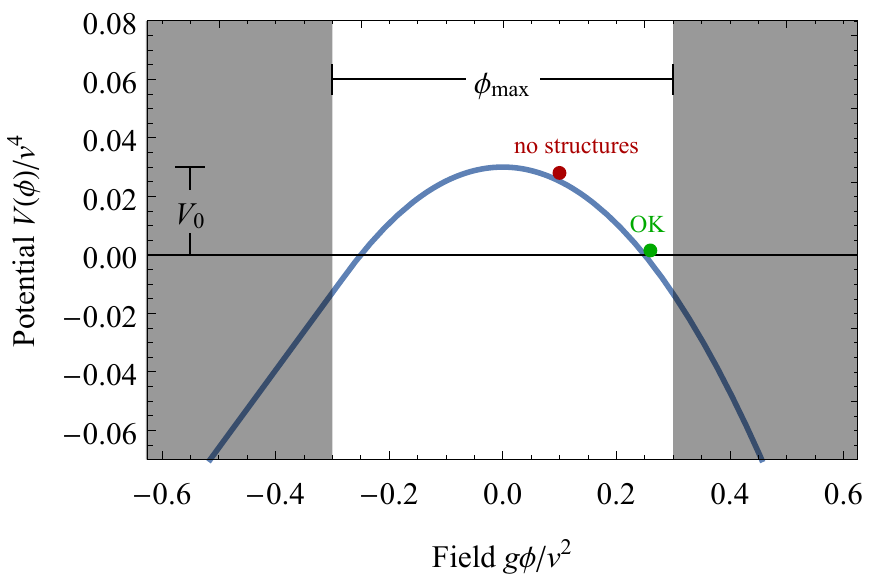} \qquad\qquad \includegraphics[width=0.4\textwidth]{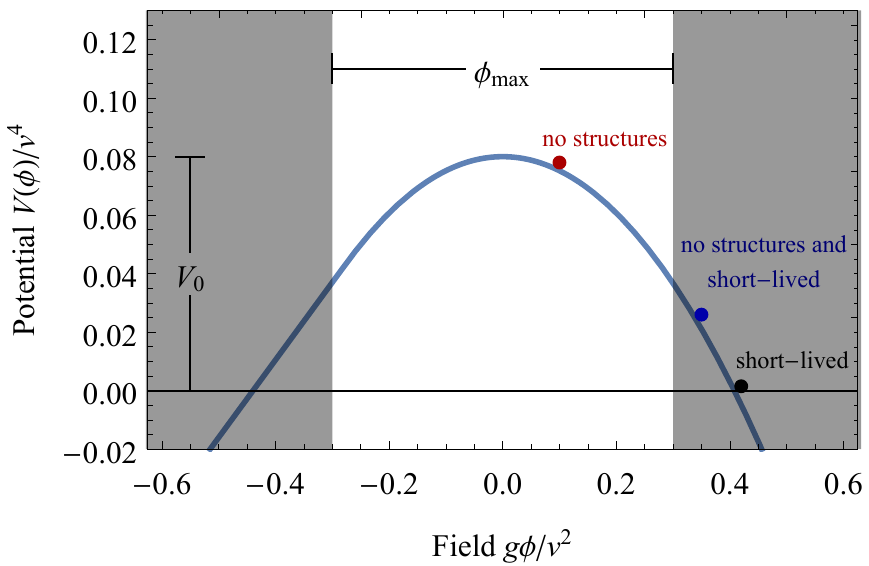}$$
\caption{\it Combined selection of a small cosmological constant, in addition to a small Higgs mass, for $m \ll H_0$. See the text for details. \label{fig:CC}}
\end{figure}

%

%

\subsection{Bound from inflationary dynamics}
We assumed that $\phi$ is near the top and nearly homogeneous within the observable horizon.
We now study if such a region can be produced by inflation with Hubble constant $\Hinf$.
The possible issue is that,
during inflation, the field $\phi$ undergoes de Sitter fluctuations $\delta \phi \approx \Hinf/{2\pi}$ per $e$-fold\footnote{Notice that the shift symmetry also suppresses non-minimal couplings of $\phi$ to gravity $f(\phi) R$, which would complicate its cosmological evolution in the early Universe.}.
About $N\approx 50$ $e$-folds of inflation are needed to produce the observable Universe from one causal patch.
During this relevant inflationary period, 
the field $\phi$ undergoes the total fluctuation $\delta \phi_{\rm infl} \approx \sqrt{N} \Hinf/2\pi $.
Such inflationary fluctuation must be smaller than $\phi_{\rm max}$ given in eq.~\eqref{eq:phirange}. 
This condition is certainly satisfied if $m\circa{<}H_0$, given that data demand $\Hinf \ll M_{\rm Pl}$.
For $m \circa{>}H_0$ we obtain the bound 
\be\label{eq:bound_infl}
m \lesssim H_0 \ln \frac{M_{\rm Pl}}{\sqrt{N} H_{\rm infl}} \sim 17 H_0\quad \hbox{for $H_{\rm infl} \sim 10^{10 } \GeV$}. 
\ee
Notice, however, that there are $e^{3N}$ patches that could have rolled down catastrophically; in the conservative approximation that their fluctuations are uncorrelated, the probability is enhanced by such $e^{3N}$ factor, modifying mildly the above bound:
\be\label{eq:bound_infl}
m \lesssim H_0 \ln \frac{M_{\rm Pl}}{N H_{\rm infl}} \sim 15 H_0\quad \hbox{for $H_{\rm infl} \sim 10^{10 } \GeV$}. 
\ee
Unlike other relaxation mechanisms, the present one is compatible with standard inflation
at mildly sub-Planckian energy.

A similar but weaker bound is obtained by considering 
fluctuations in the velocity $\dot\phi$ of the field at inflation end,
$\delta \dot\phi_{\rm infl} \sim \sqrt{3/2}\Hinf^2/2\pi$.
There is no $\sqrt{N}$ enhancement because
a free classical field satisfies $\ddot \phi + 3 (\dot a/a) \dot\phi=0$ so that its
velocity redshifts as $\dot\phi \propto 1/a^3$:
earlier inflationary  fluctuations in $\delta \dot\phi_{\rm infl}$
get diluted and can be neglected.
For the same reason an initial velocity does not lead to a large field variation in the subsequent 
post-inflationary evolution.
The total post-inflationary shift in the field is dominated by a few Hubble times
just after inflation, during which the Hubble rate still is $\approx H_{\rm infl}$ because of conservation of energy:
\beq \Delta \phi = \int dt\, \dot\phi   \approx  \frac{\delta\dot\phi_{\rm infl}}{ H_{\rm infl}} \approx\Hinf.
\eeq
This is smaller than $\delta\phi_{\rm infl}$ by a factor $\approx\sqrt{N}$ and thereby gives a weaker bound
than eq.~\eqref{eq:bound_infl}.

\subsection{Bound from thermal effects during the big-bang}
The field $\phi$ must be so weakly coupled that it negligibly thermalizes.
This is even more true at temperatures below the weak scale, where its indirect couplings to
SM particles are even more suppressed.


At temperatures above the weak scale $\med{|H|^2} = T^2/6$, so that the Lagrangian coupling $-g \phi |H|^2 /2$ becomes a thermal contribution to the $\phi$ potential,
$V_\phi^{\rm thermal} \approx - g T^2 \phi/12$, even if $g$ is very small.
Neglecting the non-thermal part of the potential, the thermal slope induces a drift of $\phi$
dominantly at the smallest temperature $T_{\rm min}\circa{>} v$ where this term is present.
In the slow-roll approximation
\be
\frac{d^2 \phi}{d t^2} + 3 H \frac{d \phi}{d t} \approx \frac{g T^2}{12}  \qquad\Rightarrow\qquad
\Delta\phi_{\rm thermal}  \approx \int \frac{g T^2}{36 H^2} \, d\ln a \sim  \frac{m M_{\rm Pl}^2}{T_{\rm min}^2}.
\ee
The rolled distance in field space $\Delta \phi_{\rm thermal}$ is within the meta-stability range $\phi_{\rm max}$ if
\be\label{eq:Troll}
m \lesssim H_0 \ln \frac{v^2}{H_0 M_{\rm Pl}} \sim 60 H_0.
\ee
The similar and stronger bound from inflationary dynamics in eq.\eq{bound_infl}
implies that the above condition is satisfied.
Thereby we do not need to study the potentially successful but more complicated
cosmological dynamics that takes place if, instead,
$\Delta \phi_{\rm thermal}$ were larger than $\phi_{\rm max}$.

 

\subsection{Bound from variations of fundamental constants}
In the present scenario the weak scale $v$ depends on $\phi$, and its variation induces
variations of other fundamental constants.
Bounds on the time variation of fundamental constants have been derived from cosmological observations.
For our purposes, the strongest bound is on the variation of
the ratio between the electron and proton mass $r = m_e /m_p$, proportional to $v$.
The bound is~\cite{cosmobound}
\beq \frac{d\ln r}{d\ln t}\circa{<}10^{-7} \qquad\hbox{at } t\sim \frac1{H_0}.\eeq
These bounds are largely subdominant 
with respect to bounds on variations of the cosmological constant.
Indeed a change in the electron mass by one part in $10^7$ implies
a huge $\approx 10^{30}$ variation in the cosmological constant.
This presumably happens in any theory
(with the possible exception of speculative theories with
self-cancellation of the cosmological constant)~\cite{hep-ph/0101130}.
In particular, this conclusion holds in the theory under consideration, where
\beq  \frac{d\ln r}{d\ln t} =\frac{d \ln v}{d\ln\phi} \times
\frac{d\ln \phi}{d\ln t} \sim  \frac{m \phi }{v^2}\times \frac{m^2}{H^2} \circa{<} \frac{m M_{\rm Pl}}{v^2} \ll 10^{-7} \,,\eeq 
having used the allowed range of $\phi$ from eq.\eq{phirange}.
In conclusion, bounds from variations of fundamental constants are negligible.

\begin{figure}[t]
$$\includegraphics[width=0.85\textwidth]{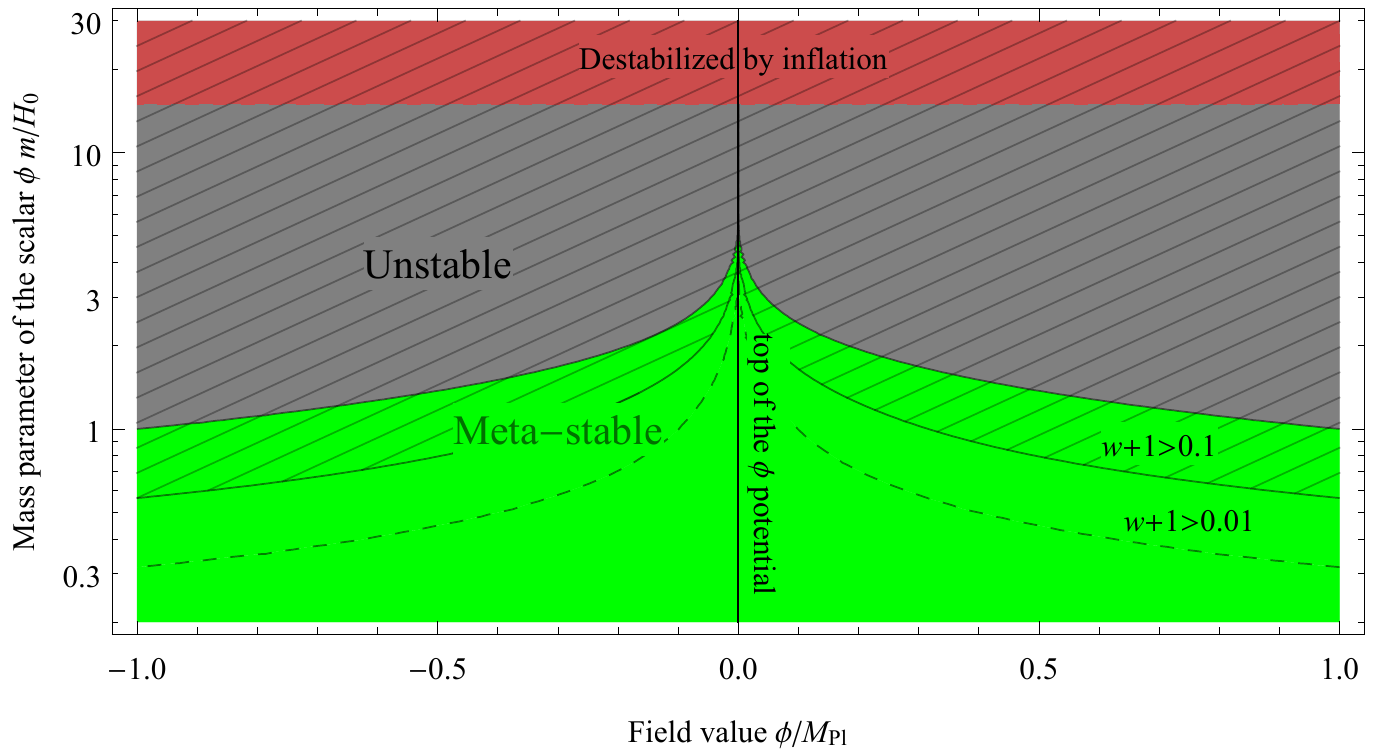}$$
\caption{\em The light scalar potential can be approximated, around its top, as $V_\phi \simeq  -m^2\phi^2/2$
so that the model depends on $m$ (vertical axis) and on the field value $\phi$ (horizontal axis).
Regions shaded in red with $m \circa{>} 15 H_0$
are excluded because inflationary perturbations are too large, in usual models
of high-scale inflation.
Regions in grey are excluded because a too fast rolling prevents a long-lived Universe.
The slightly larger hatched region is observationally 
excluded because the vacuum energy deviates too much from a cosmological constant.
Green regions are anthropically acceptable and extend down to $m \circa{>} 10^{-29} H_0$.
\label{fig:EoS}}
\end{figure}

\subsection{Signal from dark-energy equation of state}
As argued above, the only fundamental parameter that can undergo an observable cosmological
evolution is the cosmological `constant'.
Such effect is usually  parametrized by deviations from $-1$ of
its equation-of-state parameter
\beq\label{eq:w}  
w \equiv \frac{p_\phi}{\rho_\phi} = -\frac{V_\phi- K_\phi}{V_\phi+K_\phi} 
\approx -\frac{1-\phi^2/\phi_{\rm max}^2}{1+\phi^2/\phi_{\rm max}^2} ,\eeq
where $K_\phi=\dot\phi^2/2$. 
If $m \circa{>} H_0$ the red-shift of $K_\phi$ is non-negligible.
Despite this, the latter equation approximately holds because, independently from other model parameters,
the maximal field value $\phi_{\rm max}$ above which $\phi$ rolls down catastrophically fast
corresponds to order unity deviations of $w$ from $-1$.
The quadratic dependence on $\phi$  is  robust and arises because $V_\phi(\phi)$ can be approximated
as a quadratic Taylor expansion around the maximum of the potential at $\phi=0$.

The situation is illustrated in fig.\fig{EoS}.
For any given value of $m$, 
patches with  $|\phi|> \phi_{\rm max}$, corresponding to $|w+1| \circa{>} 1$, are excluded
because short-lived (grey regions).
Among the anthropically acceptable patches, an order one fraction (in hatched green)
has now been excluded by cosmological bounds.
The remaining part (in green) is fully acceptable.
This is true for both the regimes $m \circa{<} H_0$ and $m\circa{>}H_0$.\footnote{We stress here that, although at first sight figure~\ref{fig:EoS} can naively indicate that the mechanism is not effective for $m \circa{>} H_0$, 
this is not the case, since successful long-lived patches would  be generated in any case somewhere in the Universe. For $m \circa{>} H_0$ they  are simply more sparse.}
In order to discuss the expected value of $w_0$ we need to consider separately the two cases:
\begin{enumerate}

\item If $m \circa{<} H_0$ the vacuum energy significantly varies in the 
range $|\phi|<\phi_{\rm max}$.
The anthropic bound on the vacuum energy~\cite{Weinberg:1988cp}
further restricts $\phi$ to a narrower region, as illustrated in fig.\fig{CC}a.
The value of $\phi/\phi_{\rm max}$ is fixed but unknown, and determines $w_0$ as in eq.\eq{w}.

\item If $m \circa{>} H_0$ the vacuum energy does not significantly vary in the 
range $|\phi|<\phi_{\rm max}$.
The cosmological constant does not give extra restrictions on $\phi$, that is
not expected to be significantly smaller than $\phi_{\rm max}$.



\end{enumerate}
In the second case (and, perhaps, in the first case),
we expect a flat probability distribution of  $\phi$ in the range $-\phi_{\rm max}\circa{<}\phi \circa{<} \phi_{\rm max}$.
Therefore, we expect that the field $\phi$ is sliding down the potential today, 
thus having observable effects in the dark-energy equation of state.
The experimental bound $w_0 = -1.03\pm 0.03$~\cite{1807.06209} 
excludes an order unity fraction of the parameter space.
A value of $w$ within its $3\sigma$ range arises in 
a fraction $\approx \sqrt{3\times 0.03} \sim 1/3$ of the parameter space.
A more precise determination of the probability that a long-lived patch is compatible with bounds on $w$ 
requires a more detailed computation, that goes beyond the scope of this work.

In conclusion, a dynamical dark energy is a generic prediction, 
with an observable deviation of $w_0$ from $-1$ in an $\mathcal{O}(1)$ fraction of the long-lived patches.

\subsection{Collider signals}
The scenario under consideration makes the Higgs lighter than new physics by a one-loop or two-loop factor,
so that new physics is expected around the scale indicated in eq.\eq{1loop} or eq.\eq{2loop}, respectively.
Some unspecified new physics around this scale is actually needed by the scenario.
New physics could be some full solution to the hierarchy problem, such as supersymmetry.
The collider signals are well known, having been studied when natural new physics at the weak scale
was a viable possibility. 
In the one-loop case, new physics coupled to the Higgs is expected around 1.5 TeV.
This is typically consistent with current constraints from collider and precision data, as long as new physics is not colored and
does not violate flavour nor CP.  For example, in this mass range the thermal relic abundance of neutral
stable particles with electroweak interactions can reproduce the cosmological Dark Matter abundance.


\section{Conclusions}\label{sec:conclusions}
We considered an ultra-light  scalar $\phi$, tinily coupled to the Higgs and to heavier new physics. 
Its potential can easily have a maximum at field values such that the Higgs is 
significantly lighter than the new-physics scale $M$, 
since the Higgs contributes to the $\phi$ potential $V_\phi $ at tree level, while in many models
heavier new physics only contributes at one- or two-loop level.
Inflation creates different patches of the Universe with different values of $\phi$. 
Patches where $\phi$ is sufficiently close to the maximum survive for a long, cosmological, time. 
Patches with $\phi$ farther away quickly collapse into a big crunch, as seen from inside.
Therefore, a small Higgs mass is selected both anthropically (because observers can form only in long-lived regions) and environmentally (because the volume of the multiverse is dominated by long-lived regions).

\smallskip

Approximating the $\phi$  potential as $V_\phi \simeq -m^2 \phi^2/2$, the allowed range of $m$ is
\beq 10^{-29}\sim \frac{T_0^2}{v^2}\circa{<} \frac{m}{H_0} \circa{<} 10\eeq
where $T_0$ is the present temperature.
The lower bound arises demanding that the weak scale is generated. 
The upper bound arises from stability against inflationary fluctuations, considering the usual
scenario of high-scale inflation.

If $m$ is smaller than $H_0$,
the stability range of  $\phi$ covers
significantly different vacuum energies\footnote{In this case the $\phi$ field excursion is largely super-Planckian, as it is typically the case in models of relaxation.}.
The smallest $m$ allows for variations in the vacuum energy density
up to electroweak-scale ones, $v^4$.
Therefore, within this range, the model also allows for anthropic selection of the vacuum energy,
in addition to selecting a Higgs mass lighter than new physics.

The main novelty of this scenario is that the Universe is near-critical,
with $\phi$ being close to the maximum of its potential, rather than 
in a minimum. As such, naturalness is precarious: $\phi$ will necessarily roll down in the future and the
Universe will collapse into a big crunch. 
We expect that the Universe is already rolling now, with an $\mathcal{O}(1)$ probability 
to have observable effects in the dark-energy equation of state. 

Precarious naturalness introduces only a little hierarchy between the new-physics scale and the Higgs mass. 
In the most favorable case, where new physics couples to $\phi$ only via the Higgs, 
a two-loop hierarchy is generated, $M \sim (4\pi)^2 M_h \sim 20\TeV$.
More in general, 
the light scalar $\phi$ couples to the Higgs and to new physics, and
a one-loop hierarchy $M_h \sim 4\pi M_h \sim 1.5\TeV$ is generated.

%

Some special new physics, such as supersymmetry, allows for a full solution to the hierarchy problem.
However no new physics has been observed at the weak scale.
By generating a little hierarchy, the present scenario alleviates this problem, 
establishing a connection between collider  and dark-energy experiments.
These fields are presently in a precarious state, given that no new physics was discovered
despite significant efforts.
Precarious naturalness would guarantee future discoveries,
at the price of making the Universe itself precarious.
%
%
%

\footnotesize

\subsubsection*{Acknowledgments}
This work was supported by the ERC grant 669668 NEO-NAT.
We thank Alfredo Urbano and Brando Bellazzini for discussions.

\end{document}